\documentclass[12pt]{article}
\usepackage[T1]{fontenc} 
\usepackage[margin=1in]{geometry}

\usepackage{amsthm, amsmath, amssymb, mathrsfs, upgreek, enumerate, mathtools, comment, natbib, graphicx, mathabx, booktabs, threeparttable, relsize, exscale, epstopdf, scrextend, multirow, xr-hyper, pdfpages, bm, caption, subcaption, tikz, adjustbox}
\usepackage{fancyhdr}
\usepackage{easymat}
\usepackage{listings}
\usepackage{cases}
\usepackage{color}
\usepackage{pdflscape}
\usepackage{bbm}
\usepackage[section]{placeins} 

\usepackage{setspace}
\usepackage{verbatim}
\usetikzlibrary{arrows}

\usepackage[pagebackref=false]{hyperref} 
\hypersetup{
    colorlinks=true,
    citecolor=black,
    filecolor=black,
    linkcolor=black,
    urlcolor=black,
    bookmarksopen=true,
    pdfstartview=FitH
}

\newtheorem{theorem}{Theorem}
\newtheorem{assump}{Assumption}
\newtheorem{lemma}{Lemma}

\newtheorem{proposition}{Proposition}

\theoremstyle{definition}

\newtheorem{example}{Example}

\newtheorem{remark}{Remark}

\makeatletter
\def\thm@space@setup{
  \thm@preskip=15pt \thm@postskip=15pt 
}
\makeatother

\def\indep{\perp\!\!\!\perp}

\newcommand{\argmin}{\operatornamewithlimits{argmin}}

\newcommand{\E}{{\bf E}}
\newcommand{\R}{\mathbb{R}}

\newcommand{\N}{\mathcal{N}}

\newcommand{\prob}{{\bf P}}

\newcommand{\ind}{\bm{1}}
\providecommand{\abs}[1]{\lvert#1\rvert} 
\providecommand{\norm}[1]{\lVert#1\rVert}

\newcommand*{\medcap}{\mathbin{\scalebox{1.5}{\ensuremath{\cap}}}}
\newcommand*{\medcup}{\mathbin{\scalebox{1.5}{\ensuremath{\cup}}}}
\newcommand*{\medvee}{\mathbin{\scalebox{1.5}{\ensuremath{\vee}}}}

\let\emptyset\varnothing
\providecommand{\abs}[1]{\lvert#1\rvert} 
\providecommand{\norm}[1]{\lVert#1\rVert}

\renewcommand{\qed}{\hfill \mbox{\raggedright \rule{0.08in}{0.08in}}} 

\title{Neighborhood Stability in Double/Debiased Machine Learning with Dependent Data\thanks{We thank Harold Chiang for helpful discussions at an early stage of this project.}}

\author{Jianfei Cao\thanks{Department of Economics, Northeastern University. E-mail: j.cao@northeastern.edu.} \and Michael P.\ Leung\thanks{Department of Economics, University of California, Santa Cruz. E-mail: leungm@ucsc.edu.}}

\begin{document}

\maketitle
\onehalfspacing

\begin{abstract}

  {\sc Abstract.} This paper studies double/debiased machine learning (DML) methods applied to weakly dependent data. We allow observations to be situated in a general metric space that accommodates spatial and network data. Existing work implements cross-fitting by excluding from the training fold observations sufficiently close to the evaluation fold. We find in simulations that this can result in exceedingly small training fold sizes, particularly with network data. We therefore seek to establish the validity of DML without cross-fitting, building on recent work by \cite{chen2022debiased}. They study i.i.d.\ data and require the machine learner to satisfy a natural stability condition requiring insensitivity to data perturbations that resample a single observation. We extend these results to dependent data by strengthening stability to ``neighborhood stability,'' which requires insensitivity to resampling observations in any slowly growing neighborhood. We show that existing results on the stability of various machine learners can be adapted to verify neighborhood stability.

  \bigskip


 
\end{abstract}

\section{Introduction}

Double/debiased machine learning (DML) methods enable $\sqrt{n}$-consistent estimation of low-dimensional parameters, such as average treatment effects, using modern machine learning methods to approximate complex nuisance parameters. Large-sample theory for DML is mostly limited to i.i.d.\ data, but there is recent interest in applications to dependent data. \cite{brown2024inference} studies DML trained on time series data. \cite{gilbert2024causal} employ DML to control for spatial confounding. \cite{emmenegger2025treatment}, \cite{leung2025graph}, and \cite{wang2024graph} apply DML to network data.

A key theoretical challenge is verifying the high-level condition
\begin{equation}
  \left\Vert \big( M_n(\theta_0, \hat{g}) - M_n(\theta_0, g_0) \big) - \big( M(\theta_0, \hat{g}) - M(\theta_0, g_0) \big) \right\Vert = o_p(n^{-1/2}) \label{SE}
\end{equation}

\noindent where $\theta_0$ is the low-dimensional parameter of interest, $M_n$ is the empirical moment, $M$ the population moment, $g_0$ the nuisance parameter, and $\hat{g}$ the machine learning estimate of $g_0$. The goal of this paper is to provide lower-level conditions for \eqref{SE}, without cross-fitting, when the data is a $\beta$-mixing process. We allow observations to be situated in a generic metric space that can accommodate temporal, spatial, and network data. 

There are three main approaches to verifying \eqref{SE}. First, classical semiparametric estimation theory takes the approach of verifying stochastic equicontinuity, which implies \eqref{SE} \citep[e.g.][]{andrews1994asymptotics}. This condition replaces $\hat{g}$ on the left-hand side with a deterministic function $g$ and takes the supremum over all $g$ in the nuisance space $\mathcal{G}$.\footnote{Several papers verify stochastic equicontinuity for neural networks without cross-fitting. See \cite{farrell2021deep} for multilayer perceptrons and i.i.d.\ data, \cite{wang2024graph} for graph neural networks and clustered network data, and \cite{brown2024inference} for multilayer perceptrons and time series data.} As discussed in \cite{chernozhukov2018double}, this requires restrictions on the complexity of $\mathcal{G}$ that can be too strong for modern machine learning applications. \cite{chen2022debiased} note that these restrictions can be conservative since they only concern properties of $\mathcal{G}$ rather than the machine learner $\hat{g}$.

Second, we can employ cross-fitting, which essentially entails training $\hat{g}$ on a random subsample of the data (training fold) and evaluating $M_n$ on the remaining subsample (evaluation fold). When the data is i.i.d., this ensures $M_n(\theta_0, \cdot)$ and $\hat{g}$ are independent, making \eqref{SE} straightforward to verify. With dependent data, however, it is an open question how cross-fitting should be modified in general to establish \eqref{SE}. 

\cite{emmenegger2025treatment} and \cite{gilbert2024causal} consider settings with local dependence where observations are independent when separated by some minimum distance in space or a network. They naturally propose to cross-fit by excluding from the training fold observations that are sufficiently far from the evaluation fold. This modification, however, can substantially reduce the sample size of the training fold in practice, particularly with network data since the number of units at distance $k$ can grow rapidly with $k$. In the simulation study in \autoref{ssims}, the training fold size can be less than {\em ten} for reasonable parameterizations of the network formation model.

We therefore opt to avoid cross-fitting and follow a third and more recent approach due to \cite{chen2022debiased} that builds on \cite{chernozhukov2024adversarial}. They establish \eqref{SE} using a natural ``stability'' condition on the machine learner $\hat{g}$, rather than conditions on $\mathcal{G}$. Stability essentially requires $\hat{g}$ to be insensitive to perturbations of the training data that replace any observation with an independent copy. It has been commonly employed in the statistical learning literature \citep[e.g.][]{austern2020asymptotics,bayle2020cross,bousquet2002stability,celisse2016stability,elisseeff2003leave} and is advantageous due to its intuitive appeal and relative ease of verification for certain machine learners.

We show that the analysis of \cite{chen2022debiased} can be extended to dependent data. A key piece of the argument is strengthening stability to ``neighborhood'' stability. This requires insensitivity to perturbations that replace the observations in a slowly growing neighborhood of any given observation with an independent copy. Resampling neighborhoods is somewhat analogous to modifications of the bootstrap and subsampling for dependent data that resample blocks of contiguous observations. Our conditions intuitively connect the degree of dependence in the data to the strength of the restriction on the machine learner. Specifically, when the dependence between observations decays faster with distance, the neighborhood stability condition is weaker in that the machine learner needs only to be invariant to resampling smaller neighborhoods of observations.

Stability has been verified for a variety of machine learners, including decision trees \citep{arsov2019stability}, regularized M-estimators \citep{bousquet2002stability}, M-estimators trained via stochastic gradient descent \citep{hardt2016train}, and bagged learners \citep{chen2022debiased}. We show that several of these results can be adapted to verify neighborhood stability. The sufficient conditions on the machine learner are satisfied when dependence between observations at distance $r$ is small relative to the sizes of $r$-neighborhoods. This is a familiar theme for central limit theorems (CLTs) for dependent data \citep[e.g.][]{jenish2009central,kojevnikov2021limit}. When $r$-balls in Euclidean space contain more observations, a given observation is proximate to many others, so a spatial CLT requires dependence to decay more rapidly with distance to compensate.

The paper is organized as follows. The next section defines the model and estimator. Section \ref{smain} states our assumptions and the main result. We provide primitive sufficient conditions for neighborhood stability in \autoref{sns}. Section \ref{ssims} presents results from a simulation study of the \cite{emmenegger2025treatment} network interference model. Finally, \autoref{sconclude} concludes. All proofs are relegated to the appendix.

\section{Setup}\label{smodel}

Let $[n] = \{1, \ldots, n\}$ be the set of observations and $\rho_n\colon [n] \times [n] \rightarrow \R_+$ be a semi-metric, so that $\rho_n(i,j)$ represents the distance between $i,j \in [n]$. For example, observations can be located in Euclidean space with $\rho_n(i,j)$ denoting the Euclidean distance between the locations of $i$ and $j$. Alternatively, observations may be connected by a social network with $\rho_n(i,j)$ denoting the shortest path distance between $i$ and $j$. Abusing notation, for $S_1, S_2 \subseteq [n]$, let $\rho_n(S_1,S_2)=\min \{\rho_n(i,j)\colon i\in S_1, j\in S_2 \}$. For $r \in \R_+$, let $\N(i,r) = \{j \in [n]\colon \rho_n(i,j) \leq r\}$ denote the {\em $r$-neighborhood} of observation $i$. 

The data consists of the triangular array $\bm{Z} = \{Z_i\}_{i=1}^n$ where $Z_i \in \mathcal{Z} \subseteq \R^{d_z}$ for all $i \in [n]$. Let $\mathcal{G}$ be a set of functions $g\colon \mathcal{Z} \rightarrow \R^{d_g}$. We consider the moment model 
\begin{equation*}
  m(Z; \theta, g) = \psi(Z; g) \theta + \nu(Z; g)
\end{equation*}

\noindent studied, for example, by \cite{chen2022debiased} and \cite{chernozhukov2018double}. Here $\theta \in \R^p$ is the parameter of interest, $g$ a nuisance function, $\psi\colon \mathcal{Z} \times \mathcal{G} \rightarrow \R^{p\times p}$, and $\nu\colon \mathcal{Z} \times \mathcal{G} \rightarrow \R^p$. We suppose there exist ``true values'' $g_0 \in \mathcal{G}$ and $\theta_0 \in \R^p$ such that
\begin{equation*}
  \E[m(Z; \theta_0, g_0)] = 0.
\end{equation*}

\begin{example}[Average Treatment Effect]\label{eATE}
  Let $\mathcal{W}$ be a finite set of possible treatment values and $Y_i(w)$ denote the potential outcome of $i$ under the counterfactual treatment assignment $w\in\mathcal{W}$. The estimand of interest is the average treatment effect $\theta_0 = n^{-1} \sum_{i=1}^n \E[Y_i(w) - Y_i(w')]$ for some $w,w' \in \mathcal{W}$. An observation consists of $Z_i = (Y_i, W_i, X_i)$ where $W_i$ is the treatment assignment, $Y_i = Y_i(W_i)$ the outcome, and $X_i$ the vector of covariates. A popular doubly robust moment function corresponds to setting $\psi(z; g) = -1$ and
  \begin{multline*}
    \nu(z; g_0) = \frac{\bm{1}\{W_i=w\} (Y_i - \E[Y_i \mid W_i=w, X_i=x])}{\prob(W_i=w \mid X_i=x)} + \E[Y_i \mid W_i=w, X_i=x] \\ - \frac{\bm{1}\{W_i=w'\} (Y_i - \E[Y_i \mid W_i=w', X_i=x])}{\prob(W_i=w' \mid X_i=x)} - \E[Y_i \mid W_i=w', X_i=x]
  \end{multline*}

  \noindent where $x$ is the value of the covariates under $z$ \citep[][eq.\ (5.3)]{chernozhukov2018double}. Here $g_0(z) = (\E[Y_i \mid W_i=w, X_i=x], \prob(W_i=w \mid X_i=x), \E[Y_i \mid W_i=w', X_i=x], \prob(W_i=w' \mid X_i=x))$, assuming identical distributions across $i$.
\end{example}

\begin{example}[Partially Linear IV]\label{ePLIV}
  Specialize the setup of the previous example to the partially linear outcome model $Y_i = W_i\theta_0 + h_0(X_i) + \varepsilon_i$ for some unknown function $h_0$. An observation consists of $Z_i = (Y_i, W_i, X_i, V_i)$ where $V_i$ is the instrument. The Robinson-style moment function corresponds to setting
  \begin{multline*}
    \psi(z, g_0) = -(W_i - \E[W_i \mid X_i=x]) (V_i - \E[V_i \mid X_i=x]) \\ \text{and}\quad \nu(z, g_0) = (Y_i - \E[Y_i \mid X_i=x]) (V_i - \E[V_i \mid X_i=x]).
  \end{multline*}

  \noindent \citep[][eq.\ (4.8)]{chernozhukov2018double}. Here $g_0(z) = (\E[Y_i \mid X_i=x], \E[W_i \mid X_i=x], \E[V_i \mid X_i=x])$, assuming identical distributions across $i$.
\end{example}

Let $\hat{g}$ denote the machine learner, an estimate of $g_0$ trained on $\bm{Z}$. Formally, $\hat{g}$ is a mapping from $\mathcal{Z}^n$ to $\mathcal{G}$, but we suppress the dependence of $\hat{g}$ on the training set $\bm{Z}$ and simply treat $\hat{g}$ as an element of $\mathcal{G}$, writing $\hat{g}(Z_i)$. Let $\hat\theta$ be the two-step method of moments estimator that satisfies 
\begin{equation*}
  \frac1n \sum_{i=1}^n m(Z_i; \hat\theta, \hat{g}) = 0.
\end{equation*}

Our objective is to provide lower-level conditions for the following version of \eqref{SE}, which we will refer to as ``SE.'' For any $f \in \{\psi, \nu\}$, denote by $f_{k,l}$ the $kl$-th component of $f$. For the case of $f=\nu$, which is vector- rather than matrix-valued, we define $f_{k,l}=0$ whenever $k\neq l$. Let $F_{k,l}(i, g) = \E[f_{k,l}(Z_i, g)]$. SE corresponds to
\begin{multline}
  \bigg| \frac{1}{n} \sum_{i=1}^n \left( \big( f_{k,l}(Z_i, \hat{g}) - f_{k,l}(Z_i, g_0) \big) - \big( F_{k,l}(i, \hat{g}) - F_{k,l}(i, g_0) \big) \right) \bigg| = o_p(n^{-1/2}) \\\forall\, f \in \{\psi, \nu\}, \quad \{k,l\} \subseteq [p]. \label{SE2}
\end{multline}

\noindent For i.i.d.\ data, Theorem 1 of \cite{chen2022debiased} or Theorem 3.1 of \cite{chernozhukov2018double} demonstrate that $\hat\theta$ is approximately normal under SE ((A.13) of the latter reference is analogous to SE). These results can be extended to dependent data given an appropriate CLT \citep[e.g.][]{brown2024inference,gilbert2024causal,leung2025graph}, so our focus is solely on verifying SE.

\section{Main Result}\label{smain}

Let $\norm{\cdot}$ denote the Euclidean norm, and for any function $g$, let $\norm{g}_{2,i} = \E[\norm{g(Z_i)}^2]^{1/2}$.

\begin{assump}[Regularity]\label{areg}
  (a) The ranges of $\psi,\nu$ are bounded. (b) There exist constants $L,q>0$ such that $\E[(f_{k,l}(Z_i; g) - f_{k,l}(Z_i; g'))^2] \leq L\, \norm{g-g'}_{2,i}^q$ for all $f \in \{\psi,\nu\}$, $k,l \in [p]$, $g,g' \in \mathcal{G}$, $i \in [n]$, and $n\in\mathbb{N}$.
\end{assump}

\begin{example}[Average Treatment Effect]
  For the moment function in \autoref{eATE}, \autoref{areg}(a) holds if outcomes have uniformly bounded support, the propensity score $\prob(W_i=w \mid X_i)$ is bounded away from 0 and 1 for all $w$, and the machine learner $\hat{g}$ lies in a bounded set and the second component (the estimated propensity score) lies in strict subset of $(0,1)$. The latter requirement is commonly employed in the DML literature \citep[e.g.][]{chernozhukov2018double,farrell2015robust,farrell2021deep}. \autoref{areg}(b) holds for $q=2$ under the same conditions.
\end{example}

\begin{example}[Partially Linear IV]
  For the moment function in \autoref{ePLIV}, \autoref{areg}(a) holds if outcomes and instruments have uniformly bounded support and the machine learner $\hat{g}$ lies in a bounded set. \autoref{areg}(b) holds for $q=2$ under the same conditions.
\end{example}

Let $\tilde{\bm{Z}} = \{\tilde{Z}_i\}_{i=1}^n$ denote an independent copy of $\bm{Z}$ and $\hat{g}^{(-i),0}$ be the machine learner trained on $\{Z_1, \ldots, Z_{i-1}, \tilde{Z}_i, Z_{i+1}, \ldots, Z_n\}$ rather than $\bm{Z}$. The latter is a perturbed version of the data that replaces $Z_i$ with an independent copy. Let $\bm{Z}^* = \{Z_i^*\}_{i=1}^n$ denote another copy of $\bm{Z}$ independent of both $\tilde{\bm{Z}}$ and $\bm{Z}$. \cite{chen2022debiased} impose the following stability condition on the moments and $\hat{g}$ (see their Lemma 2):
\begin{multline}
  \max_{i\in [n]} \E\big[ \big( f_{k,l}(Z_i, \hat{g}) - f_{k,l}(Z_i, \hat{g}^{(-i),0}) \big)^2 \big]^{1/2} \\
  \medvee\, \max_{i \in [n]} \E\big[ \big( f_{k,l}(Z_i^*, \hat{g}) - f_{k,l}(Z_i^*, \hat{g}^{(-i),0}) \big)^2 \big]^{1/2} = o(n^{-1/2}). \label{ostab}
\end{multline}

\noindent When the machine learner is trained on i.i.d.\ data using cross-fitting, $Z_i$ and $\hat{g}$ are independent, and SE is straightforward to verify. Stability essentially weakens full independence to requiring that $Z_i$ has a negligible influence on $\hat{g}$.

To adapt \eqref{ostab} to dependent data, we replace a {\em set} of observations in a slowly growing neighborhood of $Z_i$ with an independent copy. Formally, let $\{r_n\}_{n\in\mathbb{N}} \subseteq \R_+$ be a sequence of {\bf neighborhood radii}. Construct from $\bm{Z}$ a new dataset by replacing $\{Z_k\colon k \in \N(i,r_n) \cup \N(j,r_n)\}$ with $\{\tilde{Z}_k\colon k \in \N(i,r_n) \cup \N(j,r_n)\}$, and let $\hat{g}^{(-i,-j)}$ be the machine learner trained on this data, leaving its dependence on $r_n$ implicit for economy of notation. For the case $i=j$ we abbreviate $\hat{g}^{(-i)} \equiv \hat{g}^{(-i,-j)}$.

\begin{assump}[Neighborhood Stability]\label{astable}
  There exists a sequence of neighborhood radii $\{r_n\}_{n\in\mathbb{N}}$ such that for all $f \in \{\psi, \nu\}$ and $k,l \in [p]$,
  \begin{multline*}
    \max_{i,j \in [n]} \E\big[ \big( f_{k,l}(Z_i, \hat{g}) - f_{k,l}(Z_i, \hat{g}^{(-i,-j)}) \big)^2 \big]^{1/2} \\
    \medvee\, \max_{i,j \in [n]} \E\big[ \big( f_{k,l}(Z_i^*, \hat{g}) - f_{k,l}(Z_i^*, \hat{g}^{(-i,-j)}) \big)^2 \big]^{1/2} = o(n^{-1/2}).
  \end{multline*}
\end{assump}

\noindent This strengthens \eqref{ostab} to demand invariance of $\hat{g}$ to resampling larger blocks of observations. The faster $r_n$ diverges, the stronger the requirement. How fast we require $r_n$ to diverge will depend on the degree of dependence in $\bm{Z}$, as we will see shortly. We verify \autoref{astable} for various machine learners in \autoref{sns}.

Let $\bar{N}_n(r) = n^{-1} \sum_{i=1}^n \abs{\N(i,r)}$ be the average $r$-neighborhood size. 

\begin{assump}[Learner Consistency]\label{aconsist}
  For $q$ in \autoref{areg} and $\{r_n\}_{n\in\mathbb{N}}$ in \autoref{astable}, $\max_{i \in [n]} \E[\norm{\hat{g}^{(-i)} - g_0}_{2,i}^q] = o(\bar{N}_n(r_n)^{-1})$.
\end{assump}

\noindent This requires the machine learner to converge at a sufficiently fast rate. For the case of $q=2$, with standard machine learners and weakly dependent data, we expect
\begin{equation}
  \max_{i \in [n]} \E[\norm{\hat{g} - g_0}_{2,i}^q] = o(n^{-1/2})
  \label{quarter}
\end{equation}

\noindent which corresponds to the usual $n^{-1/4}$ rate of convergence \citep[e.g.][Assumption 3.2]{chernozhukov2018double}.\footnote{Note that \eqref{quarter} contains $\hat{g}$ rather than $\hat{g}^{(-i)}$. The analogous statement for $\hat{g}^{(-i)}$ should be easier to verify since the data on which it is trained exhibits a greater degree of independence relative to $\bm{Z}$.} Then in order for \autoref{aconsist} to not be unduly restrictive, we require
\begin{equation}
  \frac{\bar{N}_n(r_n)}{\sqrt{n}} = O(1),
  \label{nbhdr}
\end{equation}

\noindent which is a restriction on the metric space and $r_n$. Choosing $r_n = O(1)$ would satisfy \eqref{nbhdr} for most spaces of interest, but our final assumption below may require $r_n$ to diverge. As we will discuss in the examples below, this together with \eqref{nbhdr} amounts to requiring sufficiently fast decay of a certain measure of dependence.

For any $\sigma$-fields $\mathcal{A},\mathcal{B}$, define the $\beta$-mixing coefficient
\begin{equation*}
  \beta(\mathcal{A}, \mathcal{B}) = \sup \frac{1}{2}\sum_{j=1}^J \sum_{k=1}^K \abs{\prob(A_j\medcap B_k) - \prob(A_j) \prob(B_k)}
\end{equation*}

\noindent where the supremum is taken over all $J,K\geq 1$ and partitions $\{A_j\}_{j=1}^J$ and $\{B_k\}_{k=1}^K$ of the sample space such that $A_j \in \mathcal{A}$ and $B_j \in \mathcal{B}$. For any $S \subseteq [n]$, let $\sigma(S)=\sigma(\{Z_i\}_{i\in S})$ be the $\sigma$-field generated by $\{Z_i\}_{i\in S}$. Define 
\begin{equation*}
  \beta_n(r) = \sup\big\{\beta(\sigma(S_1),\sigma(S_2))\colon S_1,S_2 \subseteq [n], \abs{S_1}=1, \rho_n(S_1,S_2) > r\big\}.
\end{equation*}

\noindent This measures the dependence between one observation and sets of observations at least distance $r$ away. We require sufficiently fast decay with respect to $r$ in the following sense.

\begin{assump}[Weak Dependence]\label{amix}
  For $\{r_n\}_{n\in\mathbb{N}}$ in \autoref{astable}, $n \beta_n(r_n) = o(1)$.
\end{assump}

\begin{example}[Dependence and Stability]\label{emix}
  In the commonly considered case of {\em exponential mixing} where $\sup_n \beta_n(r) \leq C e^{-\alpha r}$ for some $C, \alpha > 0$, \autoref{amix} is satisfied if $r_n = \alpha^{-1} a \log n$ for $a > 1$. This mildly strengthens standard stability conditions to resample a logarithmic ball around observations. In the case of stronger dependence with {\em polynomial mixing} where $\sup_n \beta_n(r) \leq C r^{-\alpha}$ for some $C,\alpha > 0$, the stability requirement is more demanding. Now \autoref{amix} is satisfied if $r_n = n^{a/\alpha}$ for some $a > 1$. Finally, in the case of {\em $m$-dependence}, called {\em local dependence} in the network setting \citep{chen2004normal}, units are independent after distance $m$, so $\beta_n(m) = 0$ for all $n$, and we can take $r_n = m$. Then neighborhood stability only resamples within an $m$-ball of each observation. Local dependence is a common dependence structure in settings with network interference, as discussed in \autoref{ssims}.
\end{example}

\begin{example}[Network Data]\label{enet}
  Let $\rho_n$ measure shortest path distance. As discussed in Appendix A.1 of \cite{leung2022causal}, some network formation models can feature exponential neighborhood growth where $\bar{N}_n(r) \leq C e^{\delta r}$ for some $C, \delta > 0$. We thus consider the case of exponential mixing with $r_n = \alpha^{-1} a \log n$ for $a > 1$ (\autoref{emix}). For this to satisfy \eqref{nbhdr}, we require $\alpha^{-1} a \delta \leq 0.5$. Given $a > 1$, this implies $\alpha > 2 \delta$, which says that dependence decays sufficiently quickly relative to neighborhood growth rates. Intuitively if neighborhoods grow quickly in size, then there are many observations proximate to a given ego, so weak dependence requires $\beta_n(r)$ to decay faster in $r$. As discussed in section 3.3 of \cite{leung2022causal}, the \cite{kojevnikov2021limit} CLT for network dependence requires a similar condition.
\end{example}

\begin{example}[Spatial Data]\label{espat}
   Suppose observations are located in $\R^d$ and minimally separated in space. Then $\bar{N}_n(r) \leq C r^d$ \citep[][Lemma A.1]{jenish2009central}, and we can consider the case of polynomial mixing with $r_n = n^{a/\alpha}$ for some $a > 1$ (\autoref{emix}). To satisfy \eqref{nbhdr}, we require $\alpha^{-1} ad \leq 0.5$. Given $a > 1$, this implies $\alpha > 2d$, which, as in the previous example, requires dependence to decay sufficiently quickly relative to neighborhood growth rates. The \cite{jenish2009central} CLT for spatial dependence requires an analogous condition in Assumption 3(b), which in our notation holds if $\alpha > d$.
\end{example}

\begin{remark}
  Our definition of $\beta_n(r)$ is similar to the $\beta_{1,\infty}(r)$ coefficient often used in the literature (e.g.\ (1.3) of \cite{bradley1993some} and Assumption 4(c) of \cite{jenish2009central}). The latter instead takes the supremum over sets $S_2$ of arbitrarily large size. \cite{kurisu2024gaussian} work with a similar coefficient, except they restrict $S_2$ to be a cube in $\R^d$. In their ``increasing domain'' case, $\abs{S_2}$ can grow at rate $n$, like in our definition.
\end{remark}

We now state the main result.

\begin{theorem}\label{tns}
  Under Assumptions \ref{areg}--\ref{amix}, SE \eqref{SE2} holds.
\end{theorem}

\noindent To prove the theorem, we adapt the proof of the analogous i.i.d.\ result, Lemma 2 of \cite{chen2022debiased}. We make use of a powerful $\beta$-mixing coupling lemma due to \cite{berbee1979random} (\autoref{lcoup}), which allows us to reduce certain expressions to the independent case at the cost of an approximation error involving $\beta_n$.

\section{Neighborhood Stability}\label{sns}

This section shows that existing results on the stability of machine learners can be adapted to verify neighborhood stability. We first note that \autoref{astable} is a restriction on both the moment function and machine learner. The next proposition shows that, under smoothness conditions on the moment functions, specifically a strengthened version of \autoref{areg}(b), \autoref{astable} is a consequence of a more primitive neighborhood stability condition imposed directly on the machine learner.

\begin{proposition}
  Suppose there exist $r>1$ and $L \geq 0$ such that for all $f \in \{\psi,\nu\}$, $k,l \in [p]$, $i,j \in [n]$, and $n\in\mathbb{N}$,
  \begin{align*}
    &\E\big[ \big(f_{k,l}(Z_i; \hat{g}) - f_{k,l}(Z_i; \hat{g}^{(-i,-j)})\big)^2 \big] \leq L\, \E\big[ \sup_z \norm{\hat{g}(z) - \hat{g}^{(-i,-j)}(z)}^{2r} \big]^{1/r} \quad\text{and} \\
    &\E\big[ \big(f_{k,l}(Z_i^*; \hat{g}) - f_{k,l}(Z_i^*; \hat{g}^{(-i)})\big)^2 \big] \leq L\, \E\big[ \sup_z \norm{\hat{g}(z) - \hat{g}^{(-i)}(z)}^{2r} \big]^{1/r}.
  \end{align*}

  \noindent Then \autoref{astable} holds if
  \begin{equation}
    \max_{i,j\in [n]} \E\big[ \sup_z \norm{\hat{g}(z) - \hat{g}^{(-i,-j)}(z)}^{2r} \big]^{1/(2r)} = o(n^{-1/2}).
    \label{lstable}
  \end{equation}
\end{proposition}

\noindent This is analogous to Corollary 4 of \cite{chen2022debiased}, and the proof is immediate. The remainder of this section provides primitive sufficient conditions for \eqref{lstable}.

\subsection{Regularized M-Estimators}

Suppose $\mathcal{G}$ is a reproducing kernel Hilbert space with kernel $\kappa$ and norm $\norm{\cdot}_\kappa$ \citep{berlinet2011reproducing}. Let 
\begin{equation*}
  \hat{g} = \argmin_{g\in\mathcal{G}} \left\{ \frac{1}{n}\sum_{i=1}^n \ell(g; Z_i) + \lambda \norm{g}_\kappa^2 \right\}
\end{equation*}

\noindent where $\ell\colon \mathcal{G} \times \mathcal{Z} \rightarrow \R_+$ is the loss function and $\lambda \geq 0$ the penalty parameter. This setup includes bounded SVM regression, soft margin SVM classification, and ridge regression.

Suppose all $g \in \mathcal{G}$ share a common codomain $\text{Co}(\mathcal{G})$. The following result is a simple application of Theorem 22 of \cite{bousquet2002stability}.

\begin{proposition}\label{preg}
  Suppose $\kappa$ has bounded range, $\ell$ is convex in its first argument, and $\ell$ is Lipschitz in its first argument in that there exists $\sigma \geq 0$ such that
  \begin{equation*}
    \abs{\ell(y; z) - \ell(y'; z)} \leq \sigma \abs{y - y'} \quad\forall\, y,y' \in \text{Co}(\mathcal{G}),\, z \in \mathcal{Z}.\footnote{The restrictions on $\ell$ are satisfied by quadratic loss if $\mathcal{Z}$ is bounded and $\mathcal{G}$ is totally bounded \citep{bousquet2002stability}.}
  \end{equation*}

  \noindent Then there exists $C_\kappa \geq 0$ such that for all $i,j \in [n]$ and $n\in\mathbb{N}$,
  \begin{equation*}
    \sup_z \norm{\hat{g}(z) - \hat{g}^{(-i,-j)}(z)} \leq \frac{C_\kappa^2 \sigma}{2\lambda n} \abs{\N(i,r_n) \cup \N(j,r_n)} \quad\text{a.s.}
  \end{equation*}
\end{proposition}

\noindent The result implies \eqref{lstable} when 
\begin{equation*}
  \max_{i\in[n]} \frac{\abs{\N(i, r_n)}}{\sqrt{n}} = o(\lambda \sqrt{n}).
\end{equation*}

\noindent In the standard regime where $\lambda = O(n^{-1/2})$, which results in a nondegenerate limiting distribution \citep{knight2000asymptotics}, this is similar to, though stronger than, \eqref{nbhdr}. See Examples \ref{enet} and \ref{espat}.

\subsection{Stochastic Gradient Descent}

Suppose the machine learner is parameterized by $\theta \in \Theta \subseteq \R^{d_\theta}$ so that $g(z) \equiv g(z; \theta)$, and the estimated learner is $\hat{g}(z) \equiv g(z; \hat\theta)$. To obtain $\hat\theta$, we choose an objective $\psi\colon \Theta \times \mathcal{Z} \rightarrow \R$ and seek to solve
\begin{equation*}
  \min_{\theta \in \Theta} \sum_{i=1}^n \psi(\theta; Z_i).
\end{equation*}

\noindent To compute the solution, we consider the use of stochastic gradient descent (SGD) where we initialize $\hat\theta_0 = 0$ and then apply the updating rule
\begin{equation*}
  \hat\theta_{t+1} = \hat\theta_t - \alpha_{t+1} \psi'(\hat\theta_t; Z_{t+1})
\end{equation*}

\noindent sequentially for all $t \in [n-1]$. Here $\alpha_{t+1}$ is the step size and $\psi'$ is the derivative of $\psi$ with respect to its first argument. The estimator is $\hat\theta = \hat\theta_{n-1}$. 

\cite{hardt2016train} and \cite{kissel2023black} assume $\psi$ satisifes the following smoothness and convexity conditions.

\begin{assump}[Loss Regularity]\label{aSGD} \hfill
  \begin{enumerate}[(a)]
    \item ($\gamma$-strong convexity) There exists $\gamma \geq 0$ such that for all $z \in \mathcal{Z}$ and $\theta,\theta' \in \Theta$,
      \begin{equation*}
	\psi(\theta; z) \geq \psi(\theta'; z) + \psi'(\theta'; z) (\theta-\theta') + \frac{\gamma}{2} \norm{\theta-\theta'}^2.
      \end{equation*}

    \item (Lipschitz continuity) There exist $L,\beta > 0$ such that for all $z \in \mathcal{Z}$ and $\theta,\theta' \in \Theta$,
      \begin{equation*}
	\abs{\psi(\theta; z) - \psi(\theta'; z)} \leq L\norm{\theta-\theta'} \quad\text{and}\quad \abs{\psi'(\theta; z) - \psi'(\theta'; z)} \leq \beta\norm{\theta-\theta'}.
      \end{equation*}
  \end{enumerate}
\end{assump}

Define $\hat\theta^{(-i,-j)}$ such that $\hat{g}^{(-i,-j)}(\cdot) = g(\cdot,\hat\theta^{(-i,-j)})$. The following result is a simple application of Proposition 5.1 of \cite{kissel2023black}.

\begin{proposition}\label{pSGD}
  Suppose the learning rate satisfies $\alpha_t=t^{-a}\beta^{-1}$ for some $a\in (0,1)$ such that $\gamma/\beta\geq \frac{a(1-a)}{1-2^{-(1-a)}}\frac{\log n}{n^{1-a}}$. Under \autoref{aSGD}, for all $i,j \in [n]$ and $n\in\mathbb{N}$,
  \begin{equation}
    \norm{\hat\theta - \hat\theta^{(-i,-j)}} \leq \frac{2L}{\beta} n^{-a} \abs{\N(i, r_n) \cup \N(j, r_n)} \quad\text{a.s.} \label{eSGD}
  \end{equation}
\end{proposition}

Suppose $L/\beta$ is nondegenerate. Then the right-hand side of \eqref{eSGD} is $o(n^{-1/2})$ if
\begin{equation*}
  \max_{i\in[n]} \abs{\N(i, r_n)} = o(n^{a-0.5}).
\end{equation*}

\noindent Since the left-hand side is typically non-degenerate, the learning rate $a$ must exceed 0.5. With stronger dependence ($r_n$ diverges faster) or larger neighborhood sizes ($\N(i,r)$ grows faster with $r$), the condition requires a faster learning rate $a$. In the case of $a$ close to one, the requirement is similar to, though stronger than, \eqref{nbhdr}. 

Neighborhood stability \eqref{lstable} follows under appropriate smoothness conditions on $g(\cdot; z)$, as illustrated in the next example.

\begin{example}[Ridge Regression]
  The ridge objective corresponds to
  \begin{equation*}
    \psi(\theta;z) = \frac{1}{2} (y-x'\theta)^2 + \frac{1}{2} \norm{\theta}^2
  \end{equation*}

  \noindent for $z = (y,x)$. Let $\mathbb{B}(R)$ denote the Euclidean ball with radius $R$ centered at the origin. Suppose $\mathcal{Z} = \mathbb{B}(R_z)$ and $\Theta = \mathbb{B}(R_\theta)$ for some constants $R_z,R_\theta$. Then by Example 3 of \cite{kissel2023black}, \autoref{pSGD} applies with $\gamma=\lambda$, $L=R_z^2(1+R_\theta) + \lambda R_\theta$, and $\beta=R_z^2+\lambda$, in which case
  \begin{equation*}
    \norm{g(z)-\hat{g}^{(-i,-j)}(z)} \leq \norm{x'\hat\theta - x'\hat\theta^{(-i,-j)}} \leq \frac{2(R_z^2(1+R_\theta)+\lambda R_\theta)}{R_z^2+\lambda} \frac{\abs{\N(i, r_n) \cup \N(j, r_n)}}{n^a}.
  \end{equation*}
\end{example}

\subsection{Bagged Learners}\label{sbag}

We lastly consider machine learners that average the predictions of ``base'' learners trained on resampled data, for example random forests. Following the setup of \cite{chen2022debiased}, we draw $m$ observations from $[n]$, either with or without replacement, independently $B$ times to obtain $\{\bm{Z}_{[m]}^b\}_{b=1}^B$, which we will to refer to as 	``subsamples.'' Let $\hat{h}(\cdot; \bm{Z}_{[m]}^b)\colon \mathcal{Z} \rightarrow \mathcal{G}$ be a base machine learner trained on $\bm{Z}_{[m]}^b$. The bagged learner is 
\begin{equation*}
  \hat{g}(\cdot) = \frac{1}{B} \sum_{b=1}^B \hat{h}(\cdot\,; \bm{Z}_{[m]}^{b}).
\end{equation*}

\noindent Let $\bm{Z}_{[m],(-i,-j)}^b$ be defined by replacing each element of $\{Z_k\colon k \in \N(i,r_n) \cup \N(j,r_n)\}$ in $\bm{Z}_{[m]}^b$ with the corresponding element of $\{\tilde{Z}_k\colon k \in \N(i,r_n) \cup \N(j,r_n)\}$. The following result is a simple extension of Theorem 5 of \cite{chen2022debiased}.

\begin{proposition}\label{pbag}
Let $N^*(r_n) \equiv \max_{i\in[n]} \abs{\N(i,r_n)}$. Neighborhood stability \eqref{lstable} holds for some $r>1$ under the following conditions for $s,k\geq 2r$ satisfying $s^{-1} + k^{-1} = (2r)^{-1}$.
\begin{enumerate}[(a)]
  \item (Parameters) $m N^*(r_n) = o(n^{1/2})$ and $B^{-1} \big(m N^*(r_n)\big)^{2/k} n^{1-2/k} = o(1)$. \label{abagrates}

  \item (Moments) $\E[ \sup_z \norm{\hat{h}(z, \bm{Z}_{[m]}^1)}^s ]^{1/s} \,\medvee\, \max_{i,j\in [n]} \E[ \sup_z \norm{\hat{h}(z, \bm{Z}_{[m],(-i,-j)}^1)}^s ]^{1/s} = O(1)$. \label{abagbm}
\end{enumerate}
\end{proposition}

The second part of the parameters condition holds with a sufficiently large number of subsamples $B$. The first part restricts the subsample size $m$ in a manner that depends on how fast dependence decays relative to neighborhood growth rates. The condition allows $m$ to grow polynomially with $n$ under the following strengthened version of \eqref{nbhdr}:
\begin{equation}
  \frac{N^*(r_n)}{\sqrt{n}} = O(n^{-c}) \label{nbhdr*}
\end{equation}

\noindent for some $c > 0$. Then the condition holds with $m = o(n^c)$. The strengthened condition is satisfied if the mixing rate decays fast enough relative to the neighborhood growth rate. In the case of spatial data (\autoref{espat}), this holds if $\alpha^{-1} a \delta < 0.5$. In the case of network data (\autoref{enet}), this holds if $\alpha^{-1} a d < 0.5$. 

\section{Simulation Study}\label{ssims}

We replicate the design used in \cite{emmenegger2025treatment}, which features a potential outcomes model with network interference. We find that computing their estimator without cross-fitting improves the bias and standard deviation. We also find that increasing the density of the network used in their design can result in extremely small training fold sizes when cross-fitting. 

In this design, observations are units connected through a network which is considered non-random or conditioned upon. Then $\rho_n$ is shortest path distance, and $\N(i,r)$ is the set of units is the set of units no more than path distance $r$ from $i$. For each unit we observe a scalar outcome $Y_i$, binary treatment $W_i$, covariate $C_i$, and network feature $X_i$. Outcomes are given by
\begin{equation*}
  Y_i = W_i g_1(X_i, C_i) + (1-W_i) g_0(X_i, C_i) + \varepsilon_i,
\end{equation*}

\noindent where $\E[\varepsilon_i \mid W_i, X_i, C_i] = 0$. The network features are functions of treatments and covariates of units connected to $i$, generating what the authors refer to as ``$X$-spillovers.'' Treatments are unconfounded in that $W_i \indep \varepsilon_i \mid X_i, C_i$, and the estimand of interest is the average treatment effect 
\begin{equation*}
  \theta_0 = \frac1n \sum_{i=1}^n \E[g_1(X_i,C_i) - g_0(X_i,C_i)].
\end{equation*}

\noindent We replicate the data-generating process in section 3.1 of \cite{emmenegger2025treatment} that uses an Erd\H{o}s-R\'{e}nyi network. See \autoref{ssimdet} for specifics of the model. The network formation model draws links between pairs in an i.i.d.\ fashion with linking probability $\Delta/n$ with $\Delta=3$. This means $\Delta$ is the limiting expected degree (number of connections involving a given unit), which measures network density. 

Due to the network features, the outcomes of units that share a common network neighbor are correlated since both depend upon the common neighbor's treatment. As discussed in \autoref{emix}, this induces local dependence, which satisfies $\beta$-mixing. We can take $r_n=3$ to satisfy \autoref{amix} since units are independent if they are at least path distance 3 apart and therefore lack a common neighbor.

As in \cite{emmenegger2025treatment} we estimate $\theta_0$ using the doubly robust estimator in \autoref{eATE} using random forests to approximate the propensity score and outcome regressions. We replicate the parameters of the learner specified in their section 3.1, which uses standard bootstrapped trees; see \autoref{ssimdet} for details. We also compare with subsampled trees satisfying the setup of \autoref{sbag}. Since $r_n = O(1)$, we treat $N^*(r_n)$ in \eqref{nbhdr*} as fixed and take the subsample size to be $m = \min\{10\tilde n^{1/3}, \tilde n\}$ to satisfy assumption (a) of \autoref{pbag} where $\tilde n$ is the size of the training set. 

We report results with and without cross-fitting. In the latter case, the random forests are trained on the entire dataset with the outcome regressions $\E[Y_i \mid W_i=w, X_i, C_i]$ estimated separately for $w \in \{0,1\}$. Hence $\sum_{i=1}^n W_i$ is the sample size used in the outcome regression for $w=1$. The data-generating process is calibrated such that this is about $n/2$.

Due to local dependence, \cite{emmenegger2025treatment} propose the following modification of cross-fitting to ensure independence between evaluation and training folds. First, they randomly partition $[n]$ into $K=5$ folds. For a given fold $k$ used to compute the estimator, the training fold for the random forest is the subset of the units not in fold $k$ that are not connected to a unit in $k$ and do not share a neighbor in common with any unit in $k$ (see their eq.\ (6)). The estimates are averaged across the $K$ folds to obtain $\hat\theta$. \cite{emmenegger2025treatment} further reduce the randomness of this procedure by repeating the procedure $B$ times and taking $\hat\theta$ to be the median result, but we only consider the standard case of $B=1$.

\autoref{simresults} presents the results. Whether using bootstrapped or subsampled trees, the \cite{emmenegger2025treatment} estimator sees a reduction in bias and variance when using the full sample instead of cross-fitting. We also find that bootstrapped trees perform better in terms of bias and variance, particularly when cross-fitting. Such trees correspond to a subsample size $m$ (in the notation of \autoref{pbag}) equal to $n$, which violates assumption (a) of the proposition. This indicates that, while the condition is sufficient, it may not be necessary.

\autoref{simtsizes} shows the average size of the training fold when cross-fitting for different values of $\Delta$ and $K$. We see that for the base design in \autoref{simresults}, the fold size is small, only about 10 percent of the full sample, which explains why cross-fitting results in higher variance. When we increase the expected degree to 8, the fold size becomes negligible. This is because the network is denser, and almost all units in the training fold are less than distance 3 from the evaluation fold and end up excluded from training. When the expected degree is 5 but we reduce the number of folds to two, we again obtain close to negligible sample sizes. Despite the network being sparser, half of all units are in the evaluation fold, so the vast majority of units in the training fold are again less than distance 3 from the evaluation fold.

\begin{table}[ht]
\small
\centering
\caption{Simulation Results}
\begin{threeparttable}
\begin{tabular}{llrrrrrr}
\toprule
 &  & \multicolumn{3}{c}{No Cross-fitting} & \multicolumn{3}{c}{Cross-fitting} \\
 & $n$ & 500 & 1000 & 2000 & 500 & 1000 & 2000 \\
\midrule
\multirow[c]{2}{*}{Bootstrap} & Bias & 0.0007 & 0.0005 & 0.0009 & 0.0089 & 0.0019 & 0.0008 \\
 & Std & 0.0657 & 0.0465 & 0.0331 & 0.0699 & 0.0474 & 0.0335 \\
\cmidrule(lr){2-5} \cmidrule(lr){6-8}
\multirow[c]{2}{*}{Subsampling} & Bias & 0.0092 & 0.0087 & 0.0075 & 0.0255 & 0.0219 & 0.0187 \\
 & Std & 0.0715 & 0.0499 & 0.0348 & 0.0761 & 0.0526 & 0.0366 \\
\cmidrule(lr){2-5} \cmidrule(lr){6-8}
$\sum_{i=1}^n W_i$ & & 251.69 & 503.41 & 1007.42 & 251.69 & 503.41 & 1007.42 \\
$\theta_0$ & & 0.3155 & 0.3155 & 0.3153 & 0.3155 & 0.3155 & 0.3153 \\
\bottomrule
\end{tabular}
\begin{tablenotes}[para,flushleft]
  \footnotesize 5k simulations. ``Bias'' is the absolute average difference between $\hat\theta$ and $\theta_0$ across simulation draws. ``Std'' is the standard deviation of $\hat\theta$ across simulation draws. Cross-fitting uses 5 folds.
\end{tablenotes}
\end{threeparttable}
\label{simresults}
\end{table}

\begin{table}[ht]
\small
\centering
\caption{Training Set Sizes}
\begin{threeparttable}
\begin{tabular}{lrrrrrrrrr}
\toprule
 & \multicolumn{3}{c}{$\Delta=3, K=5$} & \multicolumn{3}{c}{$\Delta=8, K=5$} & \multicolumn{3}{c}{$\Delta=5, K=2$} \\
$n$ & 500 & 1000 & 2000 & 500 & 1000 & 2000 & 500 & 1000 & 2000 \\
size & 74.12 & 148.43 & 297.06 & 0.46 & 0.95 & 1.93 & 2.03 & 4.13 & 8.25 \\
\bottomrule
\end{tabular}

\begin{tablenotes}[para,flushleft]
  \footnotesize 5k simulations. $\Delta=$ expected degree. $K=$ number of folds.
\end{tablenotes}
\end{threeparttable}
\label{simtsizes}
\end{table}

\section{Conclusion}\label{sconclude}

When cross-fitting with dependent data, existing proposals reduce correlation between training and evaluation folds by excluding from training observations too close to the evaluation set. This can result in a large number of excluded observations, particularly with network data. We establish the validity of DML methods using dependent data without cross-fitting when the machine learner is neighborhood-stable. This requires the learner to be robust to resampling data within a slowly growing neighborhood of any observation. 

When dependence is weaker in the sense that the $\beta$-mixing coefficient decays faster with distance, neighborhood stability imposes weaker restrictions on the learner in that the neighborhood can be smaller relative to the sample size. CLTs for dependent data typically require mixing coefficients to decay with distance $r$ sufficiently fast relative to the sizes of $r$-neighborhoods. We show that, under similar requirements, neighborhood stability holds for several classes of learners, including regularized $M$-estimators and bagged learners.

\appendix
\numberwithin{equation}{section} 

\section{Simulation Details}\label{ssimdet}

This section details the design and random forest parameters used in the simulation study of \cite{emmenegger2025treatment}, which we follow in \autoref{ssims}. Let $\{C_i\}_{i=1}^n \stackrel{iid}\sim \mathcal{U}([0,1])$ and $\{\varepsilon_i\}_{i=1}^n \stackrel{iid}\sim \mathcal{U}([-\sqrt{0.12}/2, \sqrt{0.12}/2])$ be independent. Treatments are drawn independently across units with $W_i \sim \text{Ber}(p(C_i))$ where
\begin{equation*}
  p(C_i) = 0.15 \cdot\ind\{C_i < 0.33\} + 0.5 \cdot\ind\{0.33 \leq C_i < 0.66\} + 0.85 \cdot\ind\{0.66 \leq C_i\}.
\end{equation*}

\noindent Let the network be drawn independently of these primitives by drawing links in an i.i.d.\ fashion across pairs with probability $\Delta/n$. Set $X_i = 0$ if $\N(i,1) = \emptyset$ and
\begin{equation*}
  X_i = \frac{1}{\abs{\N(i,1)}} \sum_{j \in \N(i,1)} (W_j + (1-W_j)) C_j
\end{equation*}

\noindent otherwise. Finally let
\begin{multline*}
  g_1(X_i, C_i) = 1.5 \cdot\ind\{X_i \in [0.5, 0.7), C_i \geq -0.2\} \\ + 4 \cdot\ind\{X_i \geq 0.7, C_i \geq -0.2\} + 0.5 \cdot\ind\{X_i \geq 0.5, C_i < -0.2\} \\ + 3.5 \cdot\ind\{X_i < 0.5, C_i \geq -0.2\} + 2.5 \cdot\ind\{X_i < 0.5, C_i < -0.2\}
\end{multline*}

\noindent and
\begin{multline*}
  g_0(X_i, C_i) = 0.5 \cdot\ind\{X_i \geq 0.4, C_i \geq 0.2\} - 0.75 \cdot\ind\{X_i \geq 0.4, C_i < 0.2\} \\ + 0.25 \cdot\ind\{X_i < 0.4, C_i \geq 0.2\} - 0.5 \cdot\ind\{X_i < 0.4, C_i < 0.2\}.
\end{multline*}

The random forests average over 500 trees. Each tree has a minimum leaf size of 5. For the propensity score, the criterion is Gini impurity, and tree depth is limited to 2.  

\section{Proofs}

Recall that $\norm{\cdot}$ is the Euclidean norm, and $\norm{g}_{2,i} = \E[\norm{g(Z_i)}^2]^{1/2}$. For any random variable $W$, let $\norm{W}_{L_q} = \E[\abs{W}^q]^{1/q}$. Recall that $\tilde{\bm{Z}}$ and $\bm{Z}^*$ are copies of $\bm{Z}$ with all three mutually independent. 

Define $\bm{Z}_{(i)} = \{Z_j\colon \rho_n(i,j) \leq r_n\}$ and $\tilde{\bm{Z}}_{(i)}$ analogously. Construct $\bm{Z}^{(-i)}$ from $\bm{Z}$ by replacing $\bm{Z}_{(i)}$ with $\tilde{\bm{Z}}_{(i)}$. Recall that $\hat{g}$ and $\hat{g}^{(-i)}$ are the machine learners trained on $\bm{Z}$ and $\bm{Z}^{(-i)}$, respectively. Let $\bm{Z}_{(-i,-j)} = \bm{Z} \backslash (\bm{Z}_{(i)} \medcup \bm{Z}_{(j)})$, and recall that $\hat{g}^{(-i,-j)}$ is the machine learner trained on $\tilde{\bm{Z}}_{(i)} \cup \tilde{\bm{Z}}_{(j)} \cup \bm{Z}_{(-i,-j)}$.

The following coupling lemma is originally due to \cite{berbee1979random}.

\begin{lemma}[\cite{viennet1997inequalities}, Lemma 5.1]\label{lcoup}
  Let $X,Y$ be random variables taking values in the Borel spaces $S_1,S_2$, respectively. Let $U \sim \mathcal{U}([0,1])$ be independent of $(X,Y)$. There exists a random variable $Y^* = f(X,Y,U)$ for a measurable function $f\colon S_1 \times S_2 \times [0,1] \rightarrow S_2$ such that $Y^* \indep X$, $Y^* \stackrel{d}= Y$, and $\prob(Y \neq Y^*) = \beta(\sigma(X), \sigma(Y))$.
\end{lemma}

\subsection{\autoref{tns}}

For economy of notation, we suppress the $k,l$ subscripts throughout the argument. It suffices to show
\begin{equation*}
  J_n \equiv \bigg|\bigg| \frac{1}{\sqrt{n}} \sum_{i=1}^n \left( \big( f(Z_i, \hat{g}) - f(Z_i, g_0) \big) - \big( F(i, \hat{g}) - F(i, g_0) \big) \right) \bigg|\bigg|_{L_1} = o(1).
\end{equation*}

\noindent As in the proof of Lemma 2 of \cite{chen2022debiased}, we break this into three components:
\begin{align*}
  J_n \leq \bigg|\bigg| \frac{1}{\sqrt{n}} \sum_{i=1}^n &\big( \big( f(Z_i, \hat{g}) - f(Z_i, \hat{g}^{(-i)}) \big) \big) \bigg|\bigg|_{L_1} + \sqrt{n} \max_{i \in [n]} \big|\big| F(i, \hat{g}) - F(i, \hat{g}^{(-i)}) \big|\big|_{L_1} \\
  &+ \bigg|\bigg| \frac{1}{\sqrt{n}} \sum_{i=1}^n \left( \big( f(Z_i, \hat{g}^{(-i)}) - f(Z_i, g_0) \big) - \big( F(i, \hat{g}^{(-i)}) - F(i, g_0) \big) \right) \bigg|\bigg|_{L_2} \\
  &\equiv J_1 + J_2 + J_3.
\end{align*}

\noindent By \autoref{astable}, 
\begin{align*}
  &J_1 \leq \sqrt{n} \max_{i \in [n]} \big|\big| f(Z_i, \hat{g}) - f(Z_i, \hat{g}^{(-i)}) \big|\big|_{L_1} = o(1) \quad\text{and} \\
  &J_2 \leq \sqrt{n} \max_{i \in [n]} \big|\big| f(Z_i^*, \hat{g}) - f(Z_i^*, \hat{g}^{(-i)}) \big|\big|_{L_1} = o(1)
\end{align*}

Turning to $J_3$, define $W_i = ( f(Z_i, \hat{g}^{(-i)}) - f(Z_i, g_0) ) - ( F(i, \hat{g}^{(-i)}) - F(i, g_0) )$, so
\begin{equation*}
  J_3^2 = \frac{1}{n} \sum_{i=1}^n \sum_{j=1}^n \E[W_i W_j] \ind\{\rho_n(i,j) \leq r_n\} + \frac{1}{n} \sum_{i=1}^n \sum_{j=1}^n \E[W_i W_j] \ind\{\rho_n(i,j) > r_n\} \equiv K_1 + K_2.
\end{equation*}

\noindent We will show that $K_1$ and $K_2$ are $o(1)$.

\paragraph{Step 1}

By \autoref{lcoup}, for all $i\in[n]$ we can construct $Z_i^*$ such that $Z_i^*\indep \{Z_j\colon j\not\in\N(i,r_n)\}$, $Z_i^* \stackrel{d}= Z_i$, and $\prob(Z_i\neq Z_i^*)\leq \beta_n(r_n)$. Define 
\begin{equation*}
  W_i^* = ( f(Z_i^*, \hat{g}^{(-i)}) - f(Z_i^*, g_0) ) - ( F(i, \hat{g}^{(-i)}) - F(i, g_0) ).
\end{equation*}

\noindent By \autoref{areg}(a), there exists $\bar{W} < \infty$ such that $\sup_n \max_i \abs{W_i} < \bar{W}$, so 
\begin{equation*}
  \E[W_i^2 - (W_i^*)^2] \leq \E[\abs{W_i^2 - (W_i^*)^2} \ind\{W_i \neq W_i^*\}] \leq 2\bar{W}^2 \prob(W_i \neq W_i^*) \leq 2\bar{W}^2 \beta_n(r_n).
\end{equation*}

Following the argument in \cite{chen2022debiased}, 
\begin{align*}
  \E\left[(W_i^*)^2\right] &\leq 2\,\E\left[(f(Z_i^*, \hat{g}^{(-i)}) - f(Z_i^*, g_0))^2\right] + 2\,\E\left[(F(i, \hat{g}^{(-i)}) - F(i, g_0))^2\right] \\
			   &= 2\,\E\left[ \E\left[(f(Z_i^*, \hat{g}^{(-i)}) - f(Z_i^*, g_0))^2 \mid \bm{Z}^{(-i)}\right] \right] + 2\,\E\left[(F(i, \hat{g}^{(-i)}) - F(i, g_0))^2\right] \\
			   &\leq 4L\, \E\left[ \norm{\hat{g}^{(-i)} - g_0}_{2,i}^q \right]
\end{align*}

\noindent where the third line uses \autoref{areg}(b). Therefore,
\begin{equation*}
  \max_{i\in[n]} \E[W_i^2] \leq 2\bar{W}^2 \beta_n(r_n) + 4L\max_{i\in[n]}  \E\left[ \norm{\hat{g}^{(-i)} - g_0}_{2,i}^q \right], 
\end{equation*}

\noindent so
\begin{equation*}
  \abs{K_1} \leq \bar{N}_n(r_n) \max_{i\in[n]} \E[W_i^2] \leq \bar{N}_n(r_n) \left( 2\bar{W}^2 \beta_n(r_n) + 4L\max_{i\in[n]} \E\left[ \norm{\hat{g}^{(-i)} - g_0}_{2,i}^q \right] \right).\footnote{It may appear that $\bar{N}_n(r_n)$ is a crude bound, particularly for larger $r_n$ growing at faster rates. Then $\rho_n(i,j) \leq r_n$ would include pairs of observations that could be fairly distant and hence weakly dependent. The issue is that the covariance between $W_i$ and $W_j$ cannot be shown to decay with $\rho_n(i,j)$ under the current assumptions. Both can depend on $\bm{Z}$ through $\hat{g}^{(-i)}$ and $\hat{g}^{(-j)}$ in a generic way, and neighborhood stability does not provide further control.}
\end{equation*}

\paragraph{Step 2}

We bound $K_2$ by adapting the ``double centering'' argument of \cite{chen2022debiased}. Define
\begin{equation*}
  W_i^{(j)} = ( f(Z_i, \hat{g}^{(-i,-j)}) - f(Z_i, g_0) ) - ( F(i, \hat{g}^{(-i,-j)}) - F(i, g_0) ).
\end{equation*}

\noindent Then
\begin{align}
  \abs{\E[W_i^{(j)}W_j]} 
  &=\big|\E\big[ W_i^{(j)} \E[W_j \mid \bm{Z}_{(i)}, \bm{Z}_{(-i,-j)}, \tilde{\bm{Z}}_{(i)}, \tilde{\bm{Z}}_{(j)}] \big]\big| \nonumber\\
  &= \big|\E\big[ W_i^{(j)} \E[W_j \mid \bm{Z}_{(i)}, \bm{Z}_{(-i,-j)}, \tilde{\bm{Z}}_{(j)}] \big]\big| \nonumber\\
  &= \big|\E\big[ W_i^{(j)} \E[W_j \mid \bm{Z}^{(-j)}] \big]\big| \nonumber\\
  &\leq \bar{W} \E\left[ \abs{\E[W_j \mid \bm{Z}^{(-j)}]} \right] \label{WW}
\end{align}

\noindent by \autoref{areg}(a). Since $\E[W_j^* \mid \bm{Z}^{(-j)}] = 0$,
\begin{equation}
  \eqref{WW} = \bar{W} \E\big[ \abs{\E[W_j-W_j^* \mid \bm{Z}^{(-j)}]} \big] \leq \bar{W} \E[\abs{W_j-W_j^*} \ind\{ Z_j\neq Z_j^*\}] \leq 2\bar{W}^2 \beta_n(r_n). \label{WW2}
\end{equation}

Let $W_i^{*(j)} = ( f(Z_i^*, \hat{g}^{(-i,-j)}) - f(Z_i^*, g_0) ) - ( F(i, \hat{g}^{(-i,-j)}) - F(i, g_0) )$. By an argument similar to \eqref{WW2},
\begin{align*}
\abs{\E[W_i^{*(j)} W_j^{(i)}]}
&= \big|\E\big[ W_i^{*(j)} \E[W_j^{(i)} \mid Z_i^*, \bm{Z}_{(-i,-j)}, \tilde{\bm{Z}}_{(i)}, \tilde{\bm{Z}}_{(j)}] \big]\big| \\
&= \big|\E\big[ W_i^{*(j)} \E[W_j^{(i)} \mid \bm{Z}_{(-i,-j)}] \big]\big| \\
&\leq 2\bar{W}^2 \beta_n(r_n),
\end{align*}

\noindent and
\begin{multline*}
  \abs{\E[W_i^{(j)} W_j^{(i)}]} = \abs{\E[W_i^{*(j)}W_j^{(i)}] + \big(\E[W_i^{(j)} W_j^{(i)}] - \E[W_i^{*(j)}W_j^{(i)}]\big)} \\
  \leq \abs{\E[W_i^{*(j)}W_j^{(i)}]} + \bar{W} \E\big[ \abs{W_i^{(j)}-W_i^{*(j)}} \big] \leq 4\bar{W}^2 \beta_n(r_n).
\end{multline*}

\noindent Combining these derivations,
\begin{multline*}
  \abs{\E[W_i W_j]} = \big| \E\big[(W_i-W_i^{(j)})(W_j-W_j^{(i)})\big] + \E[W_i^{(j)}W_j] + \E[W_j^{(i)}W_i] - \E[W_i^{(j)}W_j^{(i)}] \big| \\
  \leq \E\big[(W_i-W_i^{(j)})^2]^{1/2} \E\big[(W_j-W_j^{(i)})^2\big]^{1/2} + 8\bar{W}^2 \beta_n(r_n).
\end{multline*}

\noindent By \autoref{astable},
\begin{multline*}
  \max_{i,j\in[n]} \E\big[(W_i-W_i^{(j)})^2]^{1/2} \leq \max_{i,j\in[n]}  2\left( \E\big[(f(Z_i, \hat{g}^{(-i)}) - f(Z_i, \hat{g}^{(-i,-j)}))^2\big] \right. \\ \left. + \E\big[(f(Z_i^*, \hat{g}^{(-i)}) - f(Z_i^*, \hat{g}^{(-i,-j)}))^2\big] \right)^{1/2} = o(n^{-1/2}).
\end{multline*}

\noindent Therefore, $\abs{K_2} = O(n\beta_n(r_n))$, and 
\begin{multline*}
  \bigg|\bigg| \frac{1}{\sqrt{n}} \sum_{i=1}^n \left( \big( f(Z_i, \hat{g}) - f(Z_i, g_0) \big) - \big( F(i, \hat{g}) - F(i, g_0) \big) \right) \bigg|\bigg|_{L_2} \\ \leq \bar{N}_n(r_n) \cdot 4L\max_{i\in[n]} \E\left[ \norm{\hat{g}^{(-i)} - g_0}_{2,i}^q \right] + O(n\beta_n(r_n)).
\end{multline*}

\noindent This is $o(1)$ by Assumptions \ref{aconsist} and \ref{amix}.

\subsection{\autoref{preg}}

Fix $i,j\in[n]$, and recall the definition of $\hat{g}^{(-i),0}$ prior to \eqref{ostab}. The proof of Theorem 22 of \cite{bousquet2002stability} shows that for any $z \in \mathcal{Z}$,
\begin{equation*}
  \norm{\hat{g}(z) - \hat{g}^{(-i),0}(z)} \leq C_\kappa \norm{\hat{g} - \hat{g}^{(-i),0}}_\kappa \quad\text{and}\quad \norm{\hat{g} - \hat{g}^{(-i),0}}_\kappa \leq \frac{C_\kappa \sigma}{2\lambda n}.
\end{equation*}

\noindent This result is agnostic towards the distribution of $\bm{Z}$ and applies to our setup. Then
\begin{equation}
  \norm{\hat{g}(z) - \hat{g}^{(-i),0}(z)} \leq \frac{C_\kappa^2 \sigma}{2\lambda n}. \label{rbd}
\end{equation}

Without loss of generality, suppose the observations in $\N(i, r_n) \cup \N(j, r_n)$ are labeled $1, \ldots, m$. Let $\bm{Z}^\ell$ be the dataset obtained by replacing each element of $\{Z_k\}_{k\in[\ell]}$ with the corresponding element of $\{\tilde{Z}_k\}_{k\in [\ell]}$ and $\hat{g}^\ell$ be the analog of $\hat{g}$ trained on $\bm{Z}^\ell$. Then
\begin{equation*}
  \norm{\hat{g}(z) - \hat{g}^{(-i,-j)}(z)} \leq \sum_{\ell=1}^{m-1} \norm{\hat{g}^\ell - \hat{g}^{\ell+1}} \leq (m-1) \frac{C_\kappa^2 \sigma}{2\lambda n}
\end{equation*}

\noindent using \eqref{rbd}. \qed

\subsection{\autoref{pSGD}}

Fix $i,j\in[n]$, and let $\hat\theta^{(-i),0}$ the estimated parameters trained on $\{Z_1, \ldots, Z_{i-1}, \tilde{Z}_i, Z_{i+1}, \ldots, Z_n\}$ rather than $\bm{Z}$. Proposition 5.1 of \cite{kissel2023black}, which is agnostic towards the distribution of $\bm{Z}$, establishes \eqref{eSGD} with $\hat\theta^{(-i,-j)}$ replaced with $\hat\theta^{(-i),0}$. 

Without loss of generality, suppose the observations in $\N(i, r_n) \cup \N(j, r_n)$ are labeled $1, \ldots, m$. Let $\bm{Z}^\ell$ be the dataset obtained by replacing each element of $\{Z_k\}_{k\in[\ell]}$ with the corresponding element of $\{\tilde{Z}_k\}_{k\in [\ell]}$ and $\hat{g}^\ell$ be the analog of $\hat{g}$ trained on $\bm{Z}^\ell$. Then
\begin{equation*}
  \norm{\hat\theta - \hat\theta^{(-i,-j)}} \leq \sum_{\ell=1}^{m-1} \norm{\hat\theta^\ell - \hat\theta^{\ell+1}} \leq (m-1) \frac{2L}{\beta} n^{-a} \abs{\N(i, r_n) \cup \N(j, r_n)}.
\end{equation*}

\noindent by Proposition 5.1 of \cite{kissel2023black}. \qed

\subsection{\autoref{pbag}}

The argument closely follows the proof of Theorem 5 of \cite{chen2022debiased}. Fix $i,j \in [n]$. Let $Z_t^b$ be the $t$th observation (arbitrarily labeled) in the subsample $\bm{Z}_{[m]}^b$. Let $A_b$ be the event that $Z_t^b = Z_k$ for some $t \leq m$ and $k \in \N(i,r_n) \cup \N(j,r_n)$. Define the single-bag increment
\begin{equation*}
  \nabla\hat{h}_b(z) = \hat{h}_m(z; \bm{Z}_{[m]}^b) - \hat{h}_m(z; \bm{Z}_{[m],(-i,-j)}^b).
\end{equation*}

\noindent Then $\nabla\hat{h}_b = 0$ if $A_b$ does not occur.

Define $R_b=\sup_z \norm{\nabla\hat{h}_b(z) \ind_{A_b}} - \E[\sup_z \norm{\nabla\hat{h}_b(z) \ind_{A_b}} \mid \bm{Z}^{(+i,+j)}]$ where $\bm{Z}^{(+i,+j)} = \bm{Z} \cup \tilde{\bm{Z}}_{(i)} \cup \tilde{\bm{Z}}_{(j)}$. Then
\begin{align}
  \Big\|\sup_z\big\|\hat g(z)-\hat g^{(-i,-j)}(z)\big\|\Big\|_{L^{2r}}
  &= \left\Vert \sup_z\left\Vert \frac1B\sum_{b=1}^B \nabla\hat h_b(z)\ind_{A_b}\right\Vert\right\Vert_{L^{2r}} \nonumber\\
  &\leq \frac1B \left\Vert \sum_{b=1}^B \sup_z\, \norm{\nabla\hat h_b(z)\ind_{A_b}} \right\Vert_{L^{2r}} \nonumber\\
  &= \frac1B \left\Vert \sum_{b=1}^B \left(R_b + \E\left[\sup_z\, \norm{\nabla\hat h_b(z)\ind_{A_b}} \,\bigg|\, \bm{Z}^{(+i,+j)}\right]\right) \right\Vert_{L^{2r}} \nonumber\\
  &\leq \frac1B \left\Vert \sum_{b=1}^B R_b\right\Vert_{L^{2r}}+\frac1B\sum_{b=1}^B\left\Vert \E\left[\sup_z\, \norm{\nabla\hat h_b(z)\ind_{A_b}} \,\bigg|\, \bm{Z}^{(+i,+j)}\right] \right\Vert_{L^{2r}} \nonumber\\
  &= \frac1B \left\Vert \sum_{b=1}^B R_b\right\Vert_{L^{2r}} + \left\Vert \E\left[\sup_z\, \norm{\nabla\hat h_1(z)\ind_{A_1}} \,\bigg|\, \bm{Z}^{(+i,+j)}\right] \right\Vert_{L^{2r}} \nonumber\\
  &\equiv [P1] + [P2]. \label{pbag1}
\end{align}

\noindent The second-to-last line follows because subsamples are identically distributed across $b$ conditional on $\bm{Z}^{(+i,+j)}$.

Turning to the first term, 
\begin{multline}
  [P1] = \frac1B \E\bigg[ \E\bigg[ \bigg(\sum_{b=1}^B R_b\bigg)^{2r} \,\bigg|\, \bm{Z}^{(+i,+j)} \bigg] \bigg]^{\frac1{2r}} \\ 
  \leq \frac1B \E\left[ \big(\sqrt{B} C_{2r} \big)^{2r} \E\big[R_1^{2r} \mid \bm{Z}^{(+i,+j)}\big] \right]^{\frac1{2r}} = \frac{C_{2r}}{\sqrt{B}} \Vert R_1\Vert_{L^{2r}} \label{pbag2}
\end{multline}

\noindent for some constant $C_{2r}$ universal across $i$, $j$, and $n$. The inequality is due to the Marcinkiewicz-Zygmund inequality (see Lemma 7 of \cite{chen2022debiased}, which is (1.2) of \cite{rio2009moment}), which uses the fact that subsamples are i.i.d.\ across $b$ conditional on $\bm{Z}^{(+i,+j)}$. 

Turning to the second term, 
\begin{align}
  \left\Vert \E\left[\sup_z\, \norm{\nabla\hat h_1(z)\ind_{A_1}} \,\bigg|\, \bm{Z}^{(+i,+j)}\right] \right\Vert_{L^{2r}}
    &= \left\Vert \prob(A_1 \mid \bm{Z}^{(+i,+j)})\, \E\left[\sup_z\, \norm{\nabla\hat h_1(z)} \,\bigg|\, \bm{Z}^{(+i,+j)}, A_1\right] \right\Vert_{L^{2r}} \nonumber\\
    &= \left\Vert \prob(A_1)\, \E\left[\sup_z\, \norm{\nabla\hat h_1(z)} \,\bigg|\, \bm{Z}^{(+i,+j)}, A_1\right] \right\Vert_{L^{2r}} \nonumber\\
    &= \prob(A_1)\left\Vert \E\left[\sup_z\, \norm{\nabla\hat h_1(z)} \,\bigg|\, \bm{Z}^{(+i,+j)}, A_1\right] \right\Vert_{L^{2r}} \nonumber\\
    &\leq \prob(A_1)\E\left[ \E\left[\left(\sup_z\, \norm{\nabla\hat h_1(z)}\right)^{2r} \,\bigg|\, \bm{Z}^{(+i,+j)}, A_1\right] \right]^{\frac1{2r}} \nonumber\\
    &= \prob(A_1) \left\Vert \sup_z\, \norm{\nabla\hat h_1(z)} \right\Vert_{L^{2r}} \nonumber\\
    &\leq O(1)\cdot\prob(A_1). \label{pbag3}
\end{align}

\noindent The first equality is the law of total probability. The second uses independence of $A_1$ and $\bm{Z}^{(+i,+j)}$. The first inequality is Jensen's. The last is the moments assumption \eqref{abagbm}. 

Also observe that
\begin{align}
  \left\Vert \sup_z\, \norm{\nabla\hat h_b(z)\ind_{A_b}}\right\Vert_{L^{2r}}
    &= \E\left[ \left(\sup_z\, \norm{\nabla\hat h_b(z)}\right)^{2r}\ind_{A_b}\right]^{\frac1{2r}} \nonumber\\
    &\leq \E\left[ \left(\sup_z\, \norm{\nabla\hat h_b(z)}\right)^{s}\right]^{1/s}\E\left[ \ind_{A_b}\right]^{1/k} \nonumber\\
    &\leq O(1)\cdot\prob(A_1)^{1/k} \label{pbag5}
\end{align}

\noindent where the second line uses H\"{o}lder's inequality and the last line uses the moments assumption \eqref{abagbm}.

Combining \eqref{pbag1}--\eqref{pbag5},
\begin{align}
  \Big\|\sup_z\big\|&\hat g(z) - \hat g^{(-i,-j)}(z)\big\|\Big\|_{L^{2r}} \leq \frac{C_{2r}}{\sqrt{B}}\Vert R_1\Vert_{L^{2r}} + O(1)\cdot\prob(A_1) \nonumber\\
  &\leq \frac{C_{2r}}{\sqrt{B}}\left(\left\Vert \sup_z\, \norm{\nabla\hat h_b(z) \ind_{A_b}}\right\Vert_{L^{2r}} + \left\Vert \E\left[\sup_z\, \norm{\nabla\hat h_b(z) \ind_{A_b}} \,\bigg|\, \bm{Z}^{(+i,+j)} \right] \right\Vert_{L^{2r}}\right) + O(1)\cdot\prob(A_1) \nonumber\\
  &\leq \frac{C_{2r}}{\sqrt{B}} \left( \big(O(1)\cdot\prob(A_1)\big)^{1/k} + O(1)\cdot\prob(A_1) \right) + O(1)\cdot\prob(A_1). \label{pbag6}
\end{align}

\noindent By the union bound,
\begin{equation}
  \prob(A_b) \leq \sum_{k=1}^n \sum_{t=1}^m \prob(Z_t^b=Z_k) \ind\big\{k \in \N(i,r_n) \medcup \N(j,r_n)\big\} \leq \frac{m}{n} \,2 N^*(r_n). \label{pbag4}
\end{equation}

\noindent Then for $p_n \equiv mN^*(r_n)$,
\begin{equation*}
  \eqref{pbag6} = O\left( \frac1{\sqrt{B}} \left(\frac{p_n}{n}\right)^{1/k} + \frac1{\sqrt{B}}\frac{p_n}{n} + \frac{p_n}{n} \right).
\end{equation*}

\noindent By the rates assumption \eqref{abagrates}, $p_n = o(\sqrt{n})$ and $B^{-1}p_n^{2/k}n^{1-2/k} = o(1)$, so the right side is $o(n^{-1/2})$ as desired. \qed


\FloatBarrier
\phantomsection
\addcontentsline{toc}{section}{References}
\bibliography{DML_NS}{} 
\bibliographystyle{aer}

\end{document}